\documentclass{article}

\usepackage{PRIMEarxiv}
\usepackage{cite}
\usepackage{amsmath,amssymb,amsfonts}
\usepackage{algorithmic}
\usepackage{graphicx}
\usepackage{siunitx}
\usepackage{textcomp}
\usepackage{upgreek}
\usepackage[utf8]{inputenc} % allow utf-8 input
\usepackage[T1]{fontenc}    % use 8-bit T1 fonts
\usepackage{hyperref}       % hyperlinks
\usepackage{url}            % simple URL typesetting
\usepackage{booktabs}       % professional-quality tables
\usepackage{amsfonts}       % blackboard math symbols
\usepackage{nicefrac}       % compact symbols for 1/2, etc.
\usepackage{microtype}      % microtypography
\usepackage{lipsum}
\usepackage{fancyhdr}       % header
\usepackage{graphicx}       % graphics
\graphicspath{{media/}}     % organize your images and other figures under media/ folder

\title{Accurate measures of regional lung air volumes from chest x-rays of small animals
}

\author{
  D. W. O'Connell\textsuperscript{1}, K. S. Morgan\textsuperscript{1}, G. Ruben\textsuperscript{1}, L. C.P. Croton\textsuperscript{1}, J. A. Pollock\textsuperscript{1}, M. K. Croughan\textsuperscript{1} \\
  \textbf{E. V. McGillick\textsuperscript{2}, M. J. Wallace\textsuperscript{2}, K. J. Crossley\textsuperscript{2}, E. J. Pryor\textsuperscript{2}}\\
  \textbf{R. A. Lewis\textsuperscript{3}, S. B. Hooper\textsuperscript{2} and M. J. Kitchen\textsuperscript{1,2}}\\
\textsuperscript{1}
School of Physics and
Astronomy, Monash University, Clayton, VIC 3800, Australia\\
\textsuperscript{2} Ritchie Centre, Monash Institute of Medical Research and the Department of Obstetrics and Gynaecology,\\ Monash University, Clayton, VIC 3800, Australia\\
\textsuperscript{3} The Department of Medical Imaging, University of Saskatchewan, Canada\\ 
\texttt{email: dylan.oconnell1@monash.edu}}

\begin{document}
\maketitle

\begin{abstract}
We present a robust technique for calculating regional volume changes within the lung from X-ray radiograph sequences captured during ventilation, without the use of computed tomography (CT). This technique is based on the change in transmitted X-ray intensity that occurs for each lung region as air displaces the attenuating lung tissue. Lung air volumes calculated from X-ray intensity changes showed a strong correlation ($R^2$=0.98) against the true volumes, measured from high-resolution CT. This correlation enables us to accurately convert projected intensity data into relative changes in lung air volume. We have applied this technique to measure changes in regional lung volumes from X-ray image sequences of mechanically ventilated, recently-deceased newborn rabbits, without the use of CT. This method is suitable for biomedical research studies and shows potential for clinical application. 
\end{abstract}

% keywords can be removed

\section{Introduction}
\label{sec:introduction}
Chronic obstructive pulmonary disease (COPD) is the third leading cause of death worldwide, and the incidence is increasing~\cite{vestbo_global_2013}. A major obstacle in lowering COPD morbidity and mortality is late diagnosis~\cite{csikesz_new_2014}. Current pulmonary function tests diagnose lung function abnormalities by measuring the volume and flow rate of air as it passes in and out of the mouth. As this does not capture regional lung information, lung diseases must develop to the point where they adversely affect global respiratory function before that disease is detected, limiting the possibility of early diagnosis~\cite{csikesz_new_2014}. 
There are a wide range of imaging techniques that can aid in the diagnosis of diseases that affect soft tissue organs like the lungs. These include magnetic resonance imaging (MRI), sonography, and X-ray imaging. Sonography and MRI are widely-used techniques when imaging soft tissue, but are typically restricted to spatial resolutions of $>1$~mm. Unfortunately, this is much larger than the distal airway structures, such as the air sacs within the lung (eg. alveoli; typically on the order of 100's of microns~\cite{foster2014mouse}), where gas exchange takes place and where some pathological changes in structure occur in the presence of disease~\cite{milne2014advanced}. Additionally, sonography cannot image beneath the pleural and therefore cannot visualize the interior of the lung as lung surface acts as a perfect reflector to the signal~\cite{app10020462}. X-ray imaging can provide much higher spatial resolution compared to MRI and sonography (usually tens of microns for pre-clinical imaging~\cite{bayat2018synchrotron}), while keeping temporal resolution high and equipment and running costs low, compared to MRI~\cite{russo_handbook_2017}.

Regional lung air volume is directly affected by any lung disease that restricts air flow or changes the mechanical properties of that lung region~\cite{hogg_pathophysiology_2004, bellardine_black_relationship_2007}. The heterogeneity in severity and location of this damage is complicated to measure, as temporal patterns of ventilation, spatial variance of chest compliance (change in volume per change in pressure), and local ventilation volumes are not easily measured~\cite{weibel_lung_2017,faffe_lung_2009}. Currently, Computed Tomography (CT) is considered the gold standard for determining regional lung volumes and alveolar dimensions. However, as CT requires multiple projections (order of 1000's) it therefore has a higher associated radiation dose when compared to single projections (radiographs). CT is also dependent on stable ventilation to avoid motion artifacts~\cite{ dubsky_synchrotron-based_2012, pialat_visual_2012} and often suffers from relatively poor temporal resolution~\cite{guerrero_dynamic_2006, pan_4d-ct_2004}. Thus, the development of alternate methods to calculate lung air volume from 2D images could have significant benefit for patients.  

Recently-developed X-ray imaging techniques that utilise x-ray phase effects have provided a variety of new ways to extract lung volume information from 2D images~\cite{bayat_synchrotron_2018}. We have previously reported a method for determining changes in regional lung air volumes from 2D monochromatic phase-contrast X-ray images ~\cite{kitchen_dynamic_2008}, but this required the chest to be within a water bath to provide a reference known total sample thickness. The speckle pattern seen in 2D propagation-based phase contrast X-ray images has also been successfully analysed to determine volumes for relatively small regions of the lung~\cite{leong_measurement_2013}. However, motion blurring during rapid breathing or ventilation can wash out the phase-contrast speckles, causing a reduction in the accuracy of air volume measurements. A related approach has used high-speed imaging to track these speckle patterns throughout the breath, via velocimetry, to calculate lung expansion and infer relative lung gas volumes~\cite{fouras_altered_2012, dubsky_synchrotron-based_2012}. However, these propagation-based imaging methods require partially-coherent radiation and large object-to-detector propagation distances to generate the necessary speckle patterns. More recently, the change in X-ray transmission has been measured in a sequence of conventional X-ray attenuation images of mouse lungs~\cite{khan_simple_2021}. Comparison of X-ray transmission to CT measurements at maximum inspiration and expiration timepoints yielded a strong positive correlation. However, this only allowed for the calculation of the end-inspiratory lung gas volume. 
 
Here, we report a robust technique for measuring regional changes in lung gas volume which does not rely on phase contrast effects nor computed tomography.  This technique takes projections of transmitted X-ray intensity and utilises the Beer-Lambert law to derive lung gas volume from the intensity changes within the chest. By applying this approach to timepoints throughout the respiratory cycle, the regional changes in lung volume during respiration can be measured. 

\section{Mathematical model}

The Beer-Lambert law states that when X-ray radiation of intensity $I_0$ propagates in the $z$-direction through an object with attenuation coefficient $\mu$, then the transmitted intensity will be
\begin{equation}
\label{eq:BL}
I = I_0 e^{-\int\mu \cdot  dz}.
\end{equation}
While the soft tissue that makes up the lungs is attenuating, the air that fills the lungs is very weakly attenuating, with $\mu_{air} \approx 0$.

During lung inflation, the chest expands to allow air entry, thereby spreading the attenuating lung tissue over a larger volume. This decreases the projected thickness of that tissue ($\int dz$) seen by a ray propagating along the optical axis $z$ as the lungs fill with air. The reduction in the density of the lung tissue therefore results in an increase in the transmitted intensity of the corresponding region of the x-ray image. This expansion of the lungs is not isotropic and occurs at different rates throughout the lung, as the shape, distribution, and flow of air into the various lobes of the lung is not homogeneous. Past investigations have employed a variety of techniques to investigate the isotropic nature of alveolar inflation (e.g.~\cite{escolar_lung_2004}). However, results have been inconclusive and have indicated that some alveoli may collapse at the same time as others inflate ~\cite{kitchen_x_ray_2015,brancazio_lung_2001}.
We now desire to derive a mathematical expression for how the intensity measured through the chest will depend upon lung air volume. We start by assuming the chest is comprised of a single material with attenuation coefficient, $\mu$, attenuating a monochromatic X-ray beam. We also assume that the bone is relatively stationary during breathing sequences, and is therefore not contributing to changes in intensity within a  sufficiently large region-of-interest (ROI). 

First, we consider how changes in regional lung air volume result in changes in X-ray attenuation. For a ROI of $N$ pixels measuring the projected intensity through a part of the lungs, an increase in lung air volume will increase the transmitted intensity relative to that measured before inflation. This can be quantified by evaluating equation~(\ref{eq:BL}) at each time point. For a given pixel, we can rewrite equation~(\ref{eq:BL}) as 
\begin{equation}
    \label{eq:ffintensity}
    \ln[\overline{I}(x,y)] = -\mu{\int dz} = -\mu T(x,y),
\end{equation}
where ${\int  dz}$ evaluates the projected thickness, $T(x,y)$, of the attenuating tissue at a given time point, and $\overline{I}(x,y)$ is the flat-field-corrected intensity, $I(x,y)/I_0(x,y)$.  For a region of a digital image, containing $N$ pixels, each of area $(\Delta x)^2$, and each measuring the projected tissue thickness $T_S(x,y)$ at that location, the lung tissue volume $V*$, at a given inflation state (fixed air volume) $S$, will be the sum of the projected thicknesses across those $N$ pixels, giving 

\begin{equation}
\label{eq:bl1}
V^* =  (\Delta x)^2\Sigma_{N}  T_\mathrm{s}(x,y).
\end{equation}
Here the symbol $*$ denotes that this is the volume of tissue calculated using the equations as they are currently defined under the stated assumptions. If equation~(\ref{eq:ffintensity}) is rearranged to solve for $T(x,y)$, and substituted into equation~(\ref{eq:bl1}), we get

\begin{equation}
\label{eq:bl2}
V^* = -(\Delta x)^2\mu^{-1} \Sigma_{N} \ln[\overline{I_\mathrm{S}}(x,y)].
\end{equation}

Importantly, the observed lung tissue volume $V^*$ will only be sensitive to changes in intensity $\overline{I_S}(x,y)$ perpendicular to the optic axis --- and hence only sensitive to tissue moving transversely out of a given pixel or image region. This means the approach is not sensitive to tissue moving along the optic axis in either direction, since the tissue would then remain within that given pixel or image region. Therefore, the observed changes in intensity will only be sensitive to changes in tissue thickness perpendicular to the optical axis. Changes that can only be measured in 2D will therefore be denoted by `$\perp$'. We now consider two different states of lung inflation, with measured volumes $V_{S1}^*$ and $V_{S2}^*$, where $V_{S2}^*$ $>$ $V_{S1}^*$. Taking the difference between $V_{S2}^*$ and $V_{S1}^*$ gives the quantity
\begin{multline}
\label{eq:bl3}
\Delta V_\perp = V^*_{s2}-V^*_{s1} = (\Delta x)^2[\Sigma_{N} \mathrm{T_{S2}}(x,y) - \Sigma_{N}  \mathrm{T_{S1}}(x,y)] = \\ 
-(\Delta x)^2\mu^{-1}[\Sigma_{N} \ln(\overline{I_\mathrm{S2}}(x,y)) -\Sigma_{N} \ln(\overline{I_\mathrm{S1}}(x,y))] .
\end{multline}

\begin{figure}[b!]
\centering
\includegraphics[width=\textwidth]{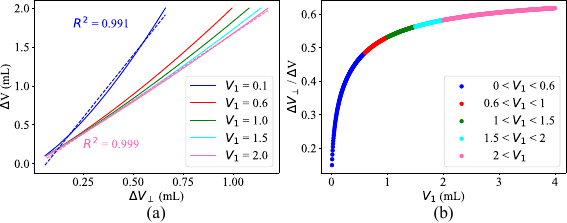}       
\caption{(a) $\Delta V_\perp$ plotted against $\Delta V$, using (\ref{G}), for typical lung volumes of small animals. A $V_1$ of 0.1~mL gives the blue curve, and applying a linear fit (dashed blue) yields $R^2 = 0.991$.  This curve also shows $\Delta V \propto \Delta V_\perp^{3/2}$. As $V_1$ increases, the trend becomes more linear, and at the highest $V_1$ plotted (2 mL), we obtain the pink curve. Again, we apply a linear fit and obtain and a $R^2 = 0.999$. (b) Here, we plot the ratio of $\Delta V_\perp$ over $\Delta V$ as a function of the initial volume $V_1$ (\ref{G}). This is helpful as it  demonstrates the typical gradients corresponding to the tissue volumes found in small animals.}
\label{fig:math_fig}
\end{figure}

Any material in the beam that has a different attenuation coefficient to the expected value (e.g. bone in soft tissue) will yield an incorrect contribution to $V_{S}^*$. However, if that object remains inside the ROI between exposures, then this incorrect volume will be subtracted to zero in the volume difference of equation~(\ref{eq:bl3}). During lung inflation, $\Delta V_\perp$, the change in the projected volume of attenuating tissue, will then be equal and opposite to the change in lung air volume ($\Delta V_\perp^{air}$), as the tissue is displaced and replaced by air (assuming the change is entirely due to additional air entering the lungs and there are no forces on the outside of the body). This means that $\Delta V_\perp^{tissue} = -\Delta V_\perp^{air}$, and hence throughout we will just consider one quantity $\Delta V_\perp$. Naturally, we desire to determine the true three-dimensional change in volume of air, $\Delta V$. 

To find the relationship between $\Delta V_\perp$ and $\Delta V$ we can consider two cubes, one with side lengths $X, Y$ and $Z$ and the second with the side lengths increased by a constant factor $a$. In an apparatus sensitive only to changes in the $x$ and $y$ direction, the change in volume between these two cubes will be
\begin{equation}
\label{vol}
    \Delta V = a^3(XYZ) - XYZ.
\end{equation}
\noindent
However, the projected change in volume can be measured from the images as
\begin{equation}
\label{perp}
    \Delta V_\perp = a^2(XYZ) - XYZ.
\end{equation}
We can rearrange equation \ref{perp} to solve for $a$ and substitute this into equation \ref{vol}, giving
\begin{equation}
\label{G}
    \Delta V= V_1 \left[\left(\frac{\Delta V_\perp}{V_1}+1\right)^{3/2} -1\right],
\end{equation}
\noindent
where $V_1=XYZ$. Equation \ref{G} implies that $\Delta V \propto \Delta V_\perp^{3/2}$. For the case of lung imaging, we measure $\Delta V_\perp$ of the tissue, where $V_1$ is the initial tissue volume. By plotting $\Delta V_\perp$ against $\Delta V$ for typical lung volumes of small animals, we obtain Figure~\ref{fig:math_fig}(a). This figure shows that when the changes in volume $\Delta V_\perp$ are small relative to the initial volume $V_1$, then $\Delta V_\perp \propto \Delta V$.

We can also see in figure~\ref{fig:math_fig}(a) the effect the initial volume $V_1$ has on the rate of change between $\Delta V_\perp$ and $\Delta V$. In figure~\ref{fig:math_fig}(b) we see, for typical ranges of tissue volumes for small animals (red, green and pink), the ratio of $\Delta V_\perp$ over $\Delta V$ will range from approximately 0.5 to 0.6. The plots of equation~(\ref{G}) shown in Figure~\ref{fig:math_fig} show that for sufficiently large initial lung tissue volumes and typical ranges of volumes for small mammals, we can approximate
\begin{equation}
\label{eq:bl4}
\Delta V \propto \Delta V_\perp. 
\end{equation}
Therefore, under the aforementioned assumptions, using equation~(\ref{eq:bl3}) to measure x-ray intensity changes, it should be possible to estimate the true change in air volume once the constants of proportionality have been determined. These constants will depend upon a given attenuation coefficient $\mu$ and given pixel size $(\Delta x)$ (see equation~(\ref{eq:bl3})) and a given range of volumes (see equation(~\ref{G})). We performed experiments to determine the validity of these assumptions and determine how accurately the volumes can hence be calculated for a given set-up.

\begin{figure}[th!]
\centering
\includegraphics[width=88mm]{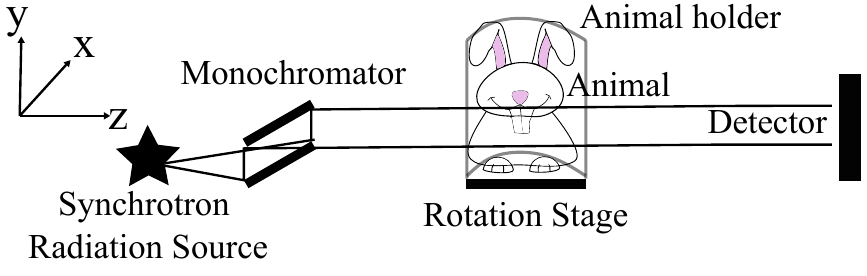}
\caption{The experimental setup used at the Imaging and Medical Beamline (IMBL) of the Australian Synchrotron. A Si(111) monochromator was used to select energies of 24~keV and 25~keV for the gated CT and breath-hold CT experiments, respectively. The deceased animals used for the experiments are held stationary using medical tape to secure them within a perspex animal holder.}
\label{fig:experiment}
\end{figure}

\section{Animal handling and ethics}
Animal experiments were approved by the Animal Ethics Committees at the Australian Synchrotron and Monash University, Australia. New Zealand white rabbit kittens were delivered by cesarean section, humanely killed via anesthetic overdose of sodium pentobarbital and intubated with an endotracheal tube inserted into the mid-cervical trachea.

\begin{figure}[th!]
\centering
\includegraphics[width=88mm]{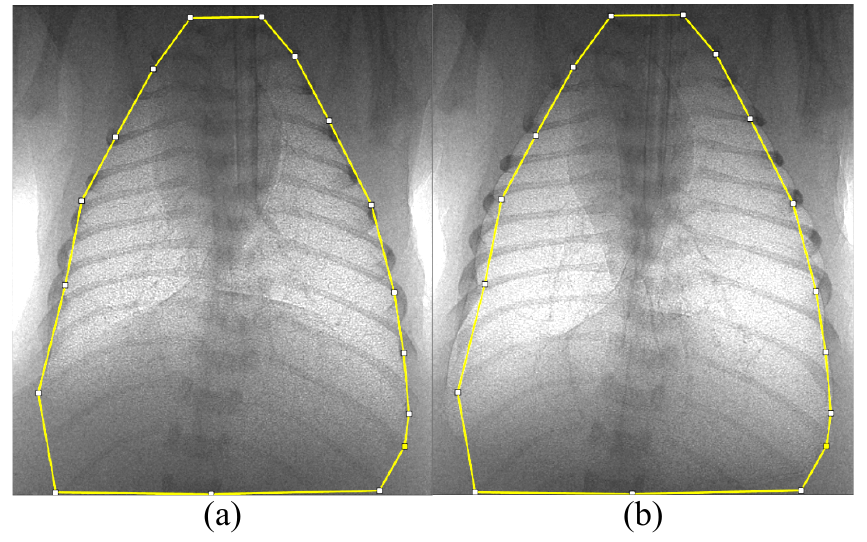}
\caption{(a) A rabbit kitten lung radiograph at lower air volume compared to (b), the highest air volume in a ventilated respiratory cycle. Both images contain a lung-shaped mask, within which the intensity can be measured. The mask is drawn around the full extent attenuating tissue. The mean intensity within the selected region increases by $\approx$ 8.5\% between (a) and (b).}
\label{fig:breaths}
\end{figure}

\begin{figure}[h!]
\centering
\includegraphics[width=88mm]{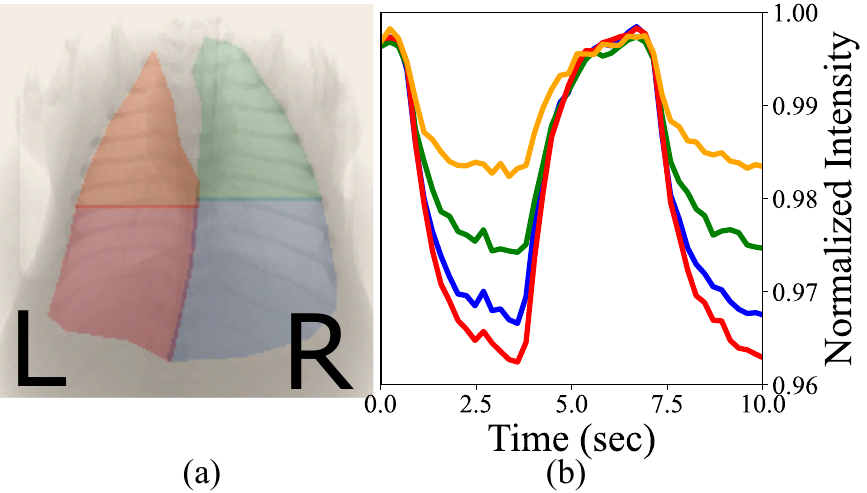}      
\caption{(a) anti-posterior projection during ventilation sequence with four regions highlighted; orange, green, red and blue. These regions correspond to the top-left, top-right, bottom-left, bottom-right, respectively. (b) The relative intensity of each region changes during ventilation of the lungs.}
\label{fig:ivstime}
\end{figure}

\section{Experimental Setup}
High-resolution images were acquired of scavenged, recently-deceased rabbit kittens at the Imaging and Medical Beamline (IMBL) of the Australian Synchrotron. The experimental setup is shown in figure~\ref{fig:experiment}. This setup makes use of the high-brilliance X-ray radiation that is monochromatized using a silicon (111) double-Laue monochromator. The imaging detector was comprised of a \SI{25}{\um}-thick Gadox phosphor (Gd\textsubscript{2}O\textsubscript{2}S:Tb; P43) coupled to a tandem lens system and a 2560$\times$2160 pixel pco.edge 5.5 sCMOS sensor, resulting in an effective pixel size of \SI{16.1}{\um}. Two CT datasets were acquired to obtain volumetric data describing the animal thoraxes in order to identify the optimum approach. A distinct difference between the two data sets is that one has the animal ventilated during the acquisition period, with projections triggered at certain breath points that can later allow for a reconstruction at each point of inflation/deflation (gated CT). Whereas the second dataset was captured with the lungs kept at a known fixed pressure during the acquisition period of the CT (breath-hold CT). The radiation energy for each approach was 24~keV and 25~keV respectively. The intensity measurements for each lung region across these sets of animals can then be compared to the corresponding volume of air extracted from the CT. The method for air volume extraction from CT data is described in Section~\ref{sec:proof}. In addition, animals were ventilated without rotation, in the antero-posterior (AP) position, with the resulting datasets used to demonstrate an application of the method, as described in Section \ref{sec:Appl}.

\section{Comparison of Regional Projected Intensity and Lung Volume Measurements}
\begin{figure}[h!]
\centering
\includegraphics[width=\textwidth]{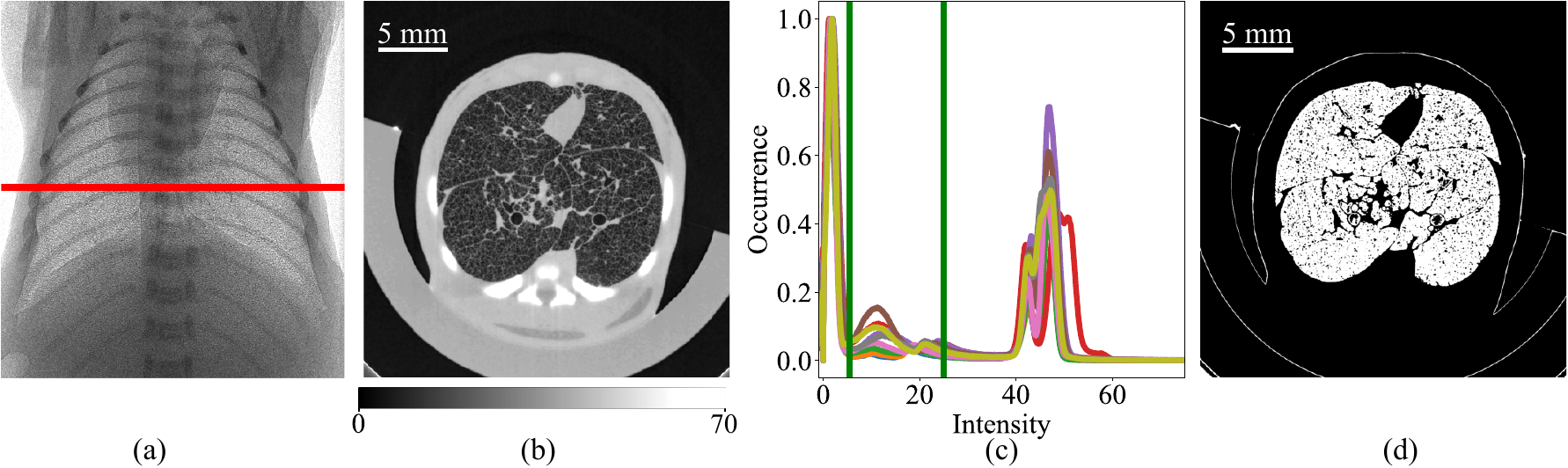}
\label{fig:flow}
\caption{(a) An AP projection of a rabbit kitten, the first from a gated CT acquisition. (b) The reconstructed CT slice at the height indicated by the red line in (a). We can represent the distribution of gray levels appearing in the CT slices, e.g. as seen in (b) using a histogram (c), given here for a set of four animals at varying pressures to demonstrate the consistency of the grey levels. After only displaying voxels in (b) with a gray level between the two green cut-off lines in (c) we obtain (d) a map of the identified air in the slice. This method falsely identifies the border of the animal holder as lung air, where the intensity drops over rapidly over a few pixels, moving down the histogram past the relevant region. However, by choosing to only consider voxels within a circular region reaching just outside the lung, we can exclude these false positives.}
\label{fig:ctflow}
\end{figure}
\label{sec:proof}

Gated computed tomography (CT) was used to obtain accurate volume measurements across a range of lung inflations. This was done by thresholding the CT dataset to isolate voxels representing air, counting those pixels, and multiplying by the voxel size.
	
In an antero-posterior (AP)  projection (or equivalently, with non-divergent X-ray setups, PA), it is straightforward to manually draw a mask around the lung tissue, or a mask just inside the rib cage, so long as the spine is sufficiently straight. Examples are provided in  figure~\ref{fig:breaths} at the maxima and minima of a ventilation cycle. The mean intensity value within the mask increases by $\approx$ 8.5\% within the mask between the maxima and minima. It is also possible to isolate sections within this region and measure how the intensity in each section changes to calculate regional lung inflation. 

Figure~\ref{fig:ivstime}(a) shows the first projection within the breath sequence, with four color-coded lung quadrants. We can record how the intensity in these regions changes as the animal is ventilated, as shown in  figure~\ref{fig:ivstime}(b). The intensity in each quadrant varies over the ventilation cycle, and each are in phase with one other. The larger lung regions have larger changes in intensity.

To calculate the reference lung volume, we use the full CT dataset, allowing us to validate the volume measurements that we aim to extract from the single projection images. Each projection can be sorted from the sequence to its associated point in the ventilation cycle to reconstruct gated CTs at each breath point. Pre-processing of each projection image before CT reconstruction included applying phase retrieval (the process of extracting phase information from images of intensity)~\cite{paganin_simultaneous_2002}, for significantly enhanced signal-to-noise ratio (SNR) with minimal loss in spatial resolution~\cite{kitchen_ct_2017}. Additionally, phase retrieval removes phase contrast effects that make segmentation difficult. To calculate the total volume within the lung from each CT, the gray values attributed to air are first isolated from a reconstructed slice. Figure~ \ref{fig:ctflow}(b) shows a representative reconstructed slice taken from the red line indicated in  figure~\ref{fig:ctflow}(a). Plotting a histogram of the gray values [figure~\ref{fig:ctflow}(c)] shows three main clusters of gray values, separated by green lines to indicate the cut-off gray values used in segmentation. The first histogram region (gray values of 0-8) is ignored as these values are attributed to regions outside the animal, where there is virtually no attenuation. We attribute the middle section to air in the lungs, where there is little attenuation. The third portion (gray values $>25$) represents the soft tissue and bones. The resolution between these sections is limited by the resolution of the detection system, as the point spread function blurs the intensity of the surrounding tissue into the airways. Figure~\ref{fig:ctflow}(d) is a binarized version of  figure~\ref{fig:ctflow}(b), subject to the condition that voxels are between the two gray values (8 and 25) indicated by the green lines in (c). This results in each CT slice being converted to a map of $N$ voxels attributed to air in the lungs. Multiplying $N$ by the voxel size ($dV$) gives the total air volume in each slice. This can be done for all slices, then summed to obtain the total lung air volume for that point in the ventilation cycle. Relative air volume, $\Delta V$, can easily be calculated by subtracting the lowest volume from the other volumes calculated at other breath points in the ventilation cycle.  

\begin{figure}[htb!]
\centering
\includegraphics[width=88mm]{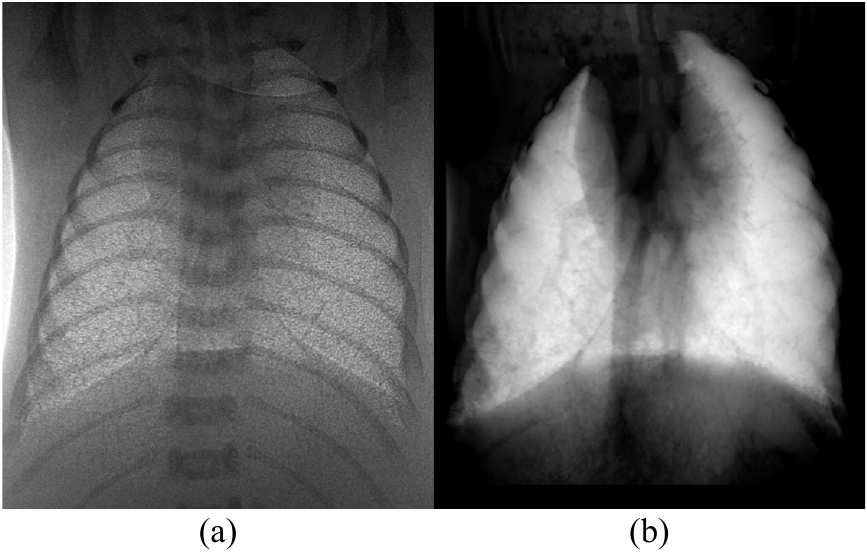} 
\caption{(a) X-ray projection image (captured as a single exposure) and (b) forward-projection of the lung air volume extracted from a CT, for a rabbit kitten in the AP projection for one matched breath-point.}
\label{fig:projectvsbackproject}
\end{figure}

For each breath point in the ventilation cycle, we can map the regional projected air thickness by forward-projecting the lung air volume (calculated via the method shown in  figure~\ref{fig:ctflow}) from the CT dataset along the optical axis. This then forms a reference against which to compare the single-projection measure $V^*$ (equation~(\ref{eq:bl2})). Figure~\ref{fig:projectvsbackproject} compares a typical projection (a), with the forward projection of the lung air volume (b), for a matched breath-point.  

\begin{figure}[tb!]
\centering
\includegraphics[width=88mm]{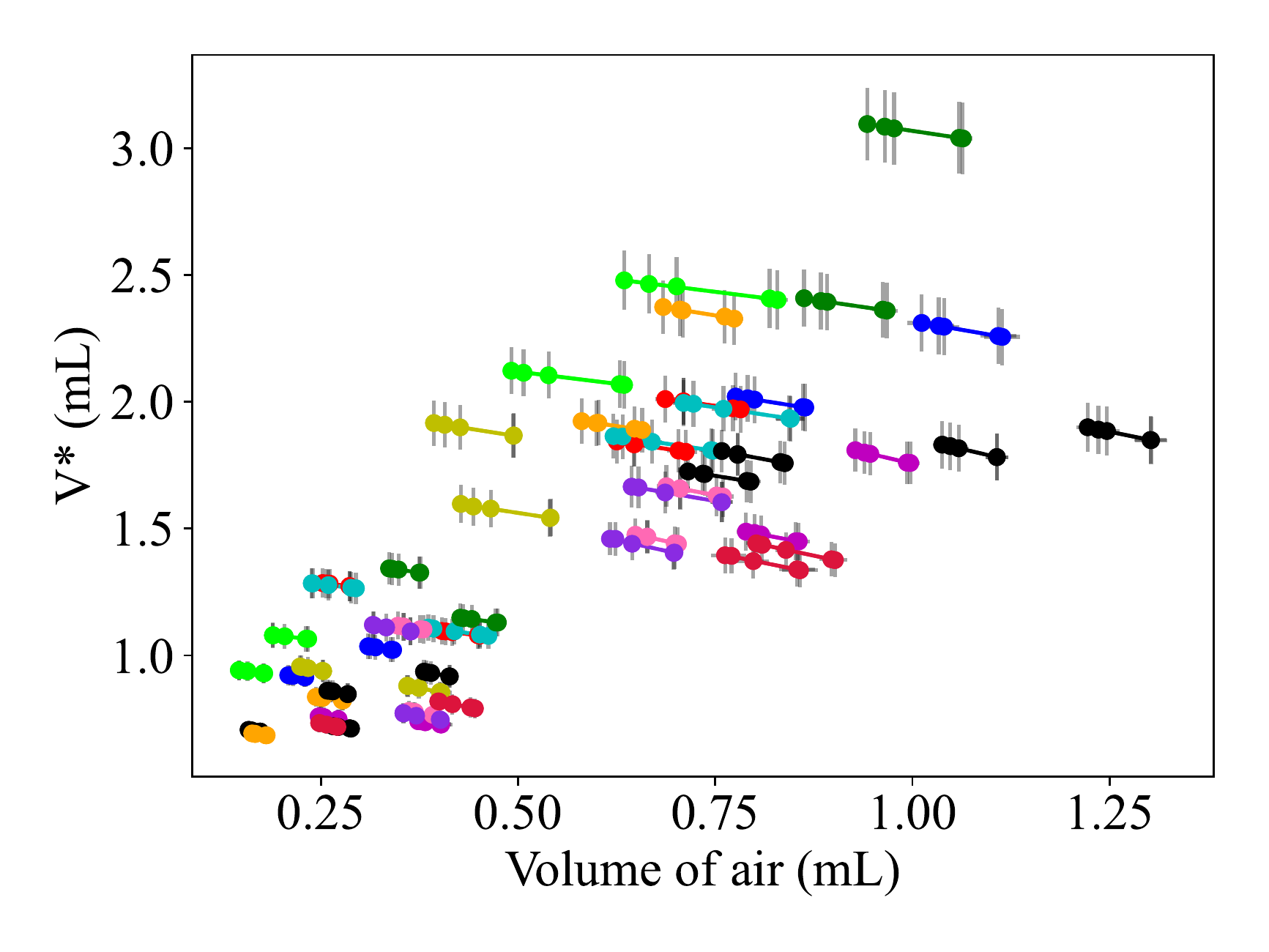}       
\caption{A plot of $V^*$, the calculated volume of tissue within the region, against the lung air volume determined by CT in the same region. Here, linear trends for each animal scan can be observed with similar gradients but different intercepts. Each color indicates a different ventilation sequence of an animal. These measurements pertain to regions sizes including lung quadrants, lung halves, and the full lung (giving multiple sequences of the same animal). Error bars were obtained by varying the projection region-of-interest sizes by $\pm~5~\%$ and recalculating the volumes.}
\label{fig:calibration}
\end{figure}

\begin{figure}[tb!]
\centering
\includegraphics[width=88mm]{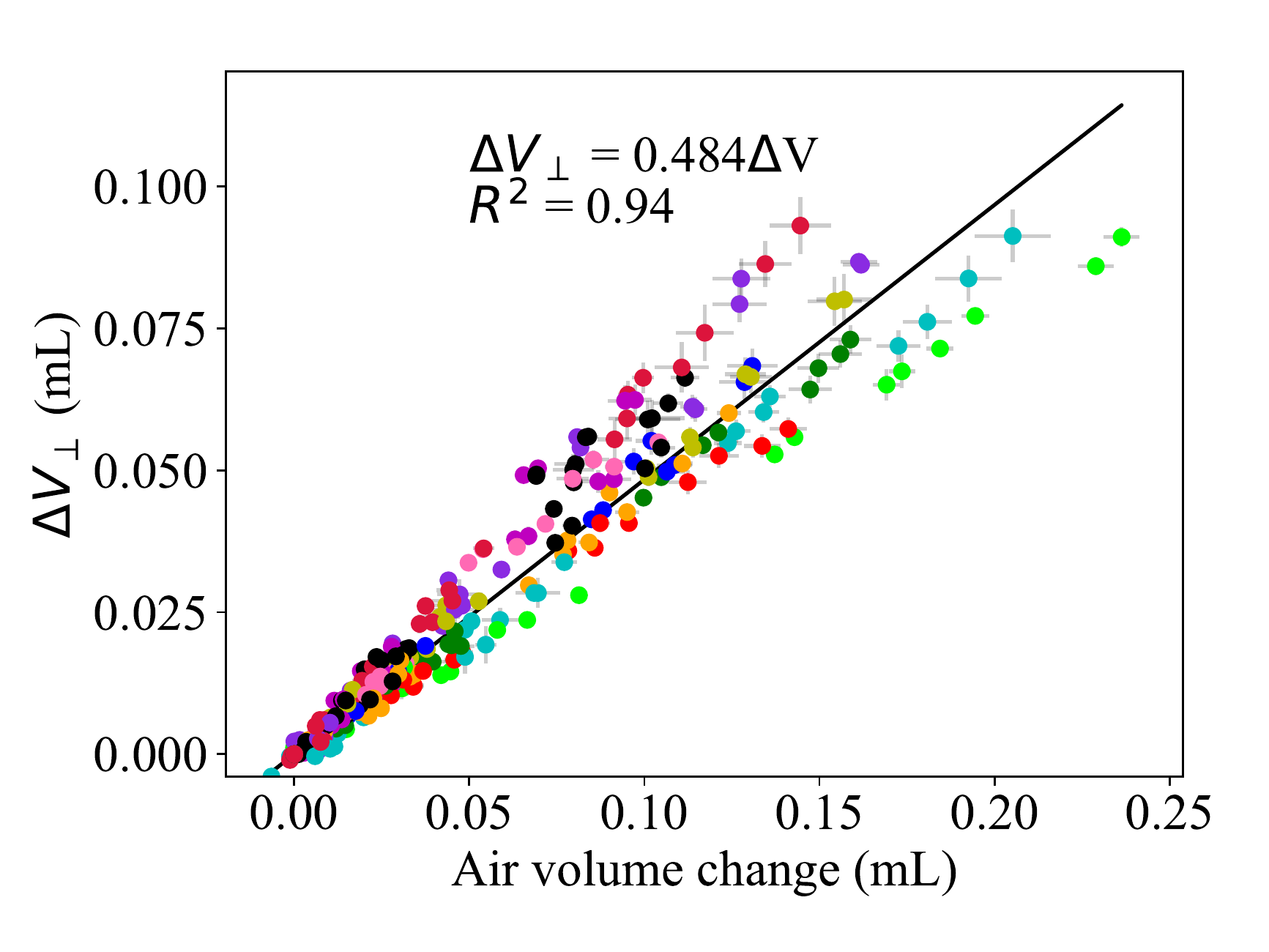}       
\caption{This plot shows $\Delta V_\perp$ vs $\Delta V$, which we obtain by subtracting the lowest volume point for each ventilation sequence observed in figure\ref{fig:calibration} from that same ventilation sequence. We then fit to a linear function ($y=mx$) that produces an $R^2$ (coefficient of determination) of 0.94 with gradient equal to $0.484 \pm 0.007$, albeit with some divergence at high lung air volumes.}
\label{fig:calibration2}
\end{figure}

Figure~\ref{fig:calibration}(a) shows the quantity $V^*$ for lung regions (quadrants, halves, and the total lung) plotted against the corresponding lung air volumes determined from CT. This is shown for eight animals for both high and low pressure scans. Gated CT often introduces motion artifacts, and scans with substantial motion artifacts are omitted. These trends are essentially a set of straight lines, one for each animal lung segment (connected by solid lines), with similar gradients but different intercepts. In order to better compare these measurements between animals, for each sequence, we can go on to subtract the $V^*$ at the lowest breath-point from each $V^*$ measurement, and similarly subtract the initial air volume for each air volume measurement. This gives $\Delta V_\perp$ and $\Delta V$ respectively. This allows us to plot a relative intensity-based measurement ($\Delta V_\perp$) vs. relative lung volume ($\Delta V$), where every point is relative to end-expiration (lowest lung volume). The result is shown in  figure~\ref{fig:calibration2}, resulting in a linear trend with a fitted gradient of $0.484 \pm 0.007$ and $R^2$ of 0.94. However, by using Gated CT, we are only able to record measurements over small volumes changes (up to 0.25~mL). In the following section, we perform this analysis using breath-hold CT to test the relationship between $\Delta V_\perp$ vs $\Delta V$ over large volume changes.

\begin{figure}[tb!]
\centering
\includegraphics[width=88mm]{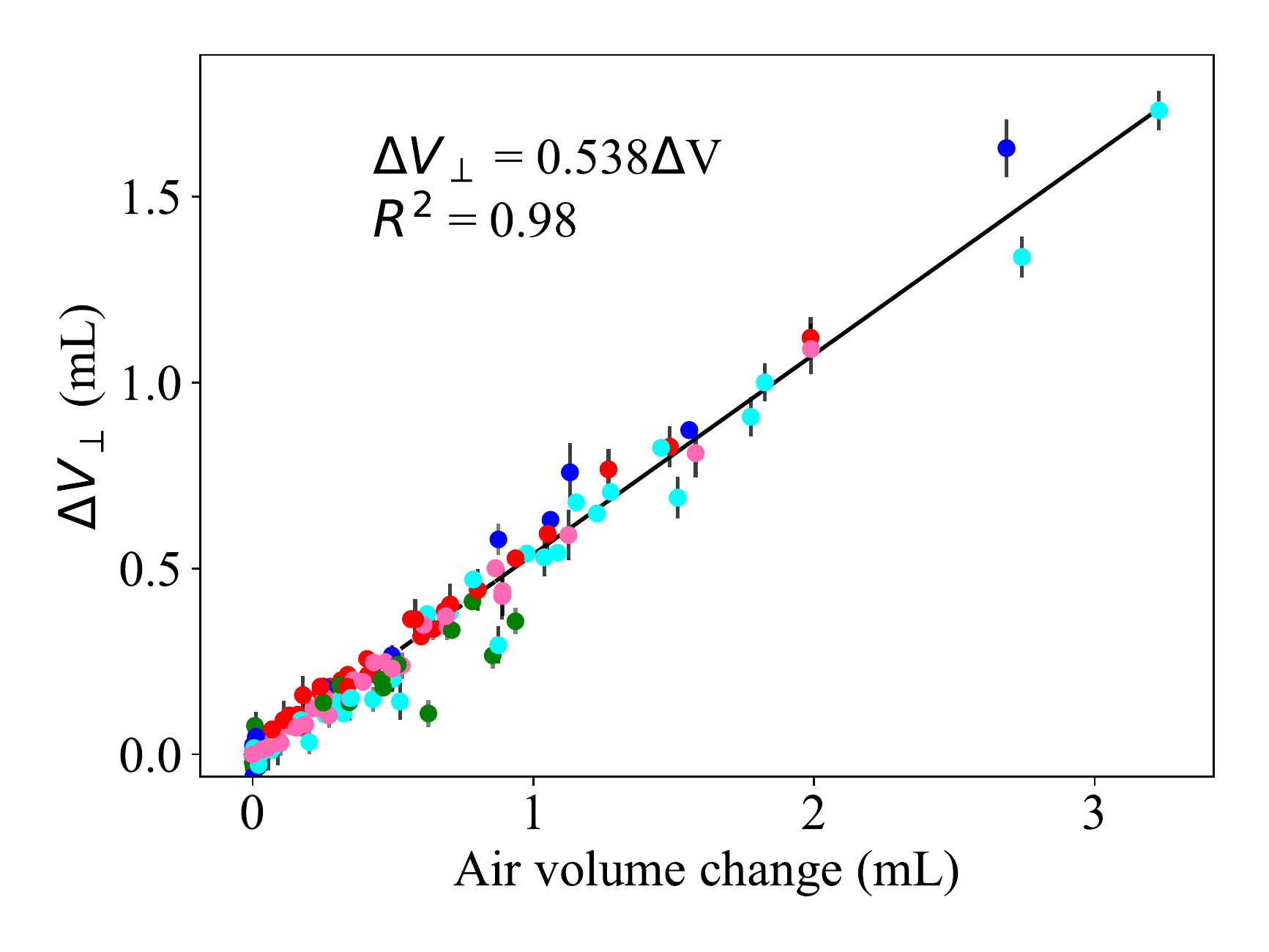}       
\caption{We plot $\Delta V_\perp$ against the change in volume of air within the corresponding region $\Delta V$ and fit to the function $y=mx$. This gives a gradient of $0.538 \pm 0.004$  fitted with an $R^2$ (coefficient of determination) of 0.98. Error bars were determined by varying mask sizes by +/- 5\%, and threshold values by +/- 10\% and taking the average of these combinations. For these plots, mask sizes of the total lung, half lung,  and quarter lung were used.}
\label{fig:IMBL_lin}
\end{figure}

\section{A large range of lung air volumes}
\label{sec:cal}

While gated CT enables us to acquire many breath sequences of a large number of animals, more stable measurements, over a wider range of volumes, were obtained by performing CT with the animal at one set airway pressure with nitrogen gas. Nitrogen was used to avoid oxygen diffusion into lung tissue and a consequent reduction in airway pressure. These were cesarean-delivered newborn animals which initially had liquid-filled lungs. Positive airway pressures were then applied to gradually increase the lung gas volume with nitrogen, finally achieving quite a large maximum relative volume of $\approx$ 3.5 mL, corresponding to an air volume to body weight ratio on the order of 100 mL/kg. In order to determine the precision of both our intensity and volume measurements, we took the average of measurements from multiple lung mask sizes (+/- 5\% area, see figure~\ref{fig:breaths}(a) for a typical mask), and we varied the range of gray level values considered to be air by +/- 10\% (i.e., shifting the vertical green lines in figure~\ref{fig:ctflow}(c)). The resulting $\Delta V_\perp$ vs. $\Delta V$ trends are shown in  figure~\ref{fig:IMBL_lin}. 

Figure~\ref{fig:IMBL_lin} shows a highly correlated linear relationship ($R^2=0.98$) between $\Delta V_\perp$ and $\Delta V$ at large lung volumes for animals of this size (newborn rabbit pups). The determined gradient compares well with the expected values for ratios of $\Delta V_\perp$ and $\Delta V$ shown in  figure~\ref{fig:math_fig}(b). Figure~\ref{fig:IMBL_lin} is a calibration plot we can use to convert any intensity change measurement---for small animals at this setup---into a volume change measurement by using the gradient of this graph ($0.504 \pm 0.006$), as described in the following section.

\section{Dynamic regional lung air volume measurements}
\label{sec:Appl}

\begin{figure}[h!]
\centering
\includegraphics[width=88mm]{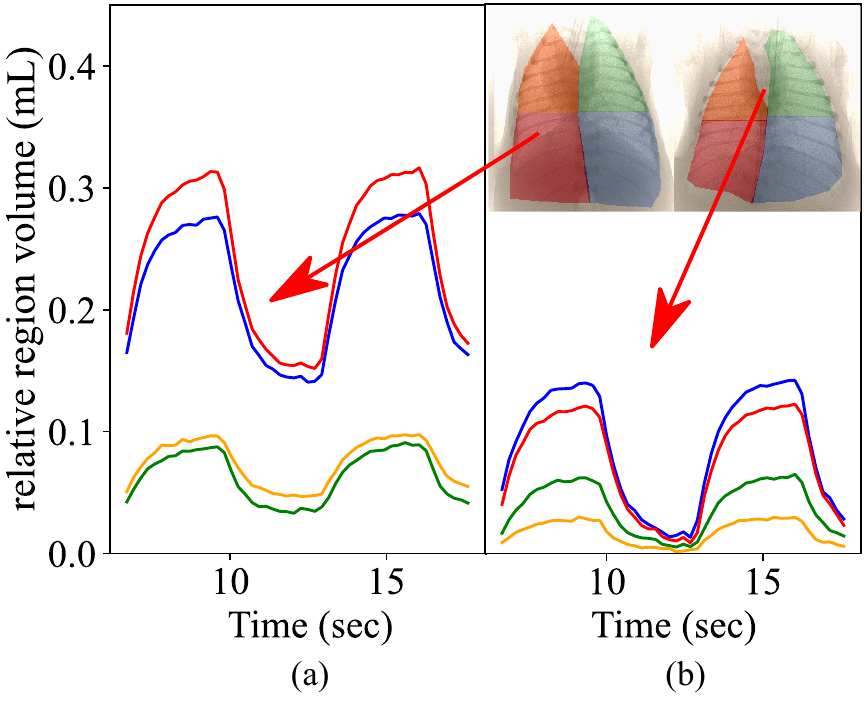}       
\caption{The relative intensity change data for two animals (for each---45 images spanning 10 seconds) was converted into relative volume measurements, shown in (a) and (b). For each animal, quadrants of the lung were measured separately, and the color-coded regions are shown in the top right for each of the two animals. Using these plots we can observe that the animal shown in (a) is getting more even aeration between each lung as the minima for each quadrant is similar when compare to the animal in (b)}
\label{fig:breath_seq}
\end{figure}

To demonstrate how this technique can be applied in dynamic imaging, the calibration (fig. ~\ref{fig:IMBL_lin}) was applied to a time sequence of X-ray AP projections of a newborn rabbit kitten being ventilated, hence the same voxel and attenuation coefficient expected as used in the calibration. With both the calibration plot for the detector (figure~\ref{fig:IMBL_lin}) and a mask for each quadrant of the lung, we determine the relative lung air volume at each breath point in the image sequence. Using this technique, we plot in vivo measurements of the volume within small regions (quadrants of the lung). Figure~\ref{fig:breath_seq} shows the change in air volume as a function of time. These plots show very minimal fluctuations in signal due to noise, showing the robustness of the technique. We note that, in figure~\ref{fig:breath_seq}(a), we have significant air volumes in the lower lobes of the lungs at all pressures with around 0.15mL of air remaining at the end of the breath. However, in figure~\ref{fig:breath_seq}(b), the lower right lobe (and upper right to a lesser extent) inflates more at high pressures that the lobes on the left side of the lung. This shows the capability of the intensity measurements to detect inhomogeneity in lung aeration, which is an important parameter for maintaining healthy lung function~\cite{hooper_respiratory_2016}.

\section{Discussion}
\label{sec:discuss}
High-resolution CT is the clinical standard for volumetric imaging of the lungs. However, this generally has a significant associated radiation dose and poor temporal resolution. The method presented in this paper requires just the change in X-ray intensity with inspiration and a calibration plot to determine the change in air volume of a region of the lung. Since we can incorporate pixel size in the measurement, the main point of specificity is the X-ray energy at which the calibration plot is generated. This approach leads to a large reduction in radiation dose and an increase in temporal resolution when compared to clinical CT. Our method measures relative air volume changes, rather than absolute volumes, since the starting intensity in a single image will vary significantly between animals of different shapes and weights. Note that this is not a limitation when considering studies where the initial image intensity corresponds to a state of no aeration, such as studies capturing the first few breaths of a newborn~\cite{wheeler_establishing_2013,te_pas_optimizing_2016}. More significance can be highlighted when comparing it to other published techniques for determining air volume within a lung of an animal without the use of CT. For example, the technique described by Leong et al.~\cite{leong_measurement_2013} has been useful for biomedical research~\cite{dekker_increasing_2019} and could, in practice, be extended to clinical use, but it relies on assumptions about the shape and packing of alveoli and even expansion throughout the lung. These assumptions may break down at high volumes, and phase contrast can be blurred during periods of rapid movement. This method also requires an X-ray source to have high spatial coherence, which is a limiting factor in clinical application. Khan et al.'s application of the observed linear relationship between X-ray intensity changes and volume changes showed similar strong correlations~\cite{khan_simple_2021} to those presented in this paper [figure~\ref{fig:IMBL_lin}(a)]. However, their approach required a high sampling rate to calculate an accumulated end-inspiration volume. Here, we are able to calculate the relative lung volume at any point in the ventilation sequence using a single calibration plot. 

To determine the limitations of this technique, we first consider the feasibility of obtaining a calibration plot. Here, we captured CT scans with voxel side lengths of \SI{15.9}{\um} and \SI{16.1}{\um}, which has allowed voxels attributed to air to be sufficiently separated from other materials in the CT data. Additionally, CTs were reconstructed using projections from a brilliant, monochromatic radiation synchrotron source. The volume accuracy is limited by the resolution of the CT system (as the minor airways must be resolvable). However, other options such for estimating lung volume from CT data, such as counting fractions of pixels as air depending on their intensity, or by morphological separation, may be able to overcome this limitation~\cite{fouras_altered_2012}. In addition, dual-energy phase contrast X-ray imaging can allow for accurate multi-material separation, which would improve air volume measurements~\cite{9133129}. Future studies will seek to apply this method using typical lab-based sources (polychromatic, low brilliance) to test the applicability in a clinical scenario, as well as in synchrotron-based experiments involving larger animals. For current experiments involving synchrotron radiation, this work can be readily applied to determine large regional volume changes to a high degree of accuracy,  which is useful for any biomedical experiments studying air volume changes~\cite{dekker_increasing_2019, pryor_improving_2020}. 

\section{Conclusion}
 We have demonstrated that the X-ray intensity changes over time in a thorax projection image correlate highly with regional lung volume changes ($R^2 = 0.97$). This proportionality allows for the calculation of regional lung volume changes in small animals imaged $in\ vivo$. In terms of diagnostic imaging, this provides an easily-implemented method for determining lung air volume changes from direct intensity images alone, enabling imaging in a dynamic setting with minimal apparatus. This technique will be particularly useful for biomedical research studies that aim to investigate air volume changes. Our experiments were conducted using monochromatic synchrotron radiation, though further investigation using conventional X-ray sources will be required to assess the suitability of this technique for clinical application. 

\section*{Funding}
This work was funded in part by the Australian Research Council’s Discovery Grant
DP170103678, the Future Fellowship Schemes under Grant
FT160100454 and Grant FT180100374, the National Health
and Medical Research Council (NHMRC) Project under Grant 1102564, the Veski Victorian Post-Doctoral Research Fellowship (VPRF), the German Excellence Initiative, the European Union
Seventh Framework Program under Grant 291763, the International Synchrotron Access Program (ISAP) managed by the Australian
Synchrotron, a part of the Australian Nuclear Science and Technology
Organisation (ANSTO), and the Australian Government under
Grant AS/IA173/14218. Erin V. McGillick was supported by a NHMRC Peter Doherty Biomedical Early Career Fellowship (APP1138049). Dylan W. O'Connell was supported by the Faculty of Science Dean’s Postgraduate Research Scholarship awarded by the School of Physics and Astronomy at Monash University and the Research Training Program (RTP) Scholarship

\bibliographystyle{unsrt}  
\bibliography{Manuscript}

\end{document}